\documentclass[8pt]{article}
\usepackage{}
\usepackage{cite}
\usepackage[letterpaper,hmargin=1in,vmargin=1in]{geometry}
\usepackage{times}
\usepackage{multirow}
\usepackage{graphicx}
\usepackage{subfigure}
\usepackage{ae}
\usepackage{amsmath,amsthm,amssymb,amsfonts,boxedminipage}
\usepackage{bbm}
\usepackage{longtable}
\newcommand\address[1]{\hskip2.25pc \parbox{.8\textwidth}{ \noindent%
   \footnotesize \it \begin{center} #1 \end{center}\rm }  \normalsize \vskip-.2cm }

\renewcommand\title[1]{\bf \hskip2.25pc \parbox{.8\textwidth}{ \noindent%
   \LARGE \bf \begin{center} #1 \end{center} \rm } \vskip.1in \rm\normalsize }
\newcommand{\upcite}[1]{\textsuperscript{\textsuperscript{\cite{#1}}}}
\renewcommand\author[1]{\hskip2.25pc \parbox{.8\textwidth}{ \noindent%
   \normalsize \bf \begin{center} #1 \end{center}\rm } \vskip-1pc }
\makeatletter
\thm@headfont{\sc}
\makeatother
\newtheorem{athm}{\bf \t}

\begin{document}

\title{Controlled Alternate Quantum Walks based Quantum Hash Function}
\author{Dan Li$^{1,\ast}$, Yu-Guang Yang$^{2}$, Jing-Lin Bi$^{2}$, Jia-Bin Yuan $^{1}$, \& Juan Xu $^{1}$}

\address{$^{1}$ College of Computer Science and Technology, Nanjing University of Aeronautics and Astronautics, Nanjing, China\\
$^{2}$ Faculty of Information Technology, Beijing University of Technology, Beijing 100124, China}
\renewcommand{\thefootnote}{}
\footnotetext{$^{\ast}$Email: lidansusu007@163.com}

\begin{quote}
Through introducing controlled alternative quantum walks, we present controlled alternate quantum walks (CAQW) based quantum hash function. CAQW based quantum hash function have excellent security, outstanding statistical performance and splendid expansibility. Furthermore, due to the structure of alternative quantum walks, implementing CAQW based quantum hash function significantly reduces the resources necessary for its feasible experimental realization than implementing other quantum hash functions. Besides, CAQW based quantum hash function has expansibility.
\end{quote}

Hash function, as an important part of modern cryptography, has widely applications in message authentication, signature protocol and Public Key Infrastructure. There are many theoretical studies about classical hash functions and mature hash functions such as MD5, SHA1 and SHA512. These hash functions are generally constructed based on mathematics complexity and thus they are computationally secure. Now classical hash functions are used extensively in quantum protocols to ensure the safety of quantum communication\upcite{A04}, yet it is contradictory with the fact that the quantum computer could solve difficult mathematical problems. For the reason above, Li et. al.  presented a kind of quantum hash function based on two-particle controlled interacting quantum walks (CIQW)\upcite{A01,A02}. This quantum hash function guarantees the security of hash function by infinite possibilities of the initial state and the irreversibility of measurement  rather than hard mathematic problems. In Ref. \cite{A03}, Yang et. al. improved the CIQW based quantum hash function and found its applications in the privacy amplification process of quantum key distribution, pseudorandom number generation and image encryption. But this quantum hash function still have two unsatisfactory defects. One of that is for the quantum hash function \upcite{A01,A02}, predictable collisions happens for some special states as initial coin state.  The second one is that  realization of controlled two-particle interacting quantum walks needs more resource than one-particle quantum walks. Therefore, we present the quantum hash function based on controlled alternate quantum walks (CAQW) here to avoid these defects.

Quantum walk, one famous quantum computation model, has widely applications in quantum computation and quantum information\upcite{A10,A11,A12,A13,A14,A15,A16}. Alternate quantum walk is a kind of quantum walk which attracts a lot of attentions \upcite{B01,B02,B03,B04,B05,B06,B07,B08}. Di Franco et. al.\upcite{B01,B02} proposed a two-dimensional quantum walk where the requirement of a higher dimensionality of the coin space is substituted with the alternance of the directions in which the walker can move. They also proposed the N-dimensional alternate quantum walk and discussed some of its properties through the analysis of the dispersion relation\upcite{B03}. Then, the authors found\upcite{B04} that by properly choosing the measurement basis, the $x-y$ spatial entanglement can be increased with respect to the value obtained with the measurement on the computational basis. They also studied the localizationlike effect in two-dimensional alternate quantum walk with periodic coin operations\upcite{B05}. In Ref. \cite{B06}, Machida and Chandrashekar presented a three-state alternate quantum walk on a two-dimensional lattice and discussed its localization and limit laws. Recently, Bru et. al. considered the two-dimensional alternate quantum walk on a cylinder\upcite{B07}. Chen and Zhang revealed the quantum and classical behaviors of the two-dimensional alternative quantum walk in the presence of decoherence\upcite{B08}.
\vspace{3mm}
\\
\noindent\textbf{Results}

\newtheorem{definition}{\indent Definition}
\newtheorem{lemma}{\indent Lemma}
\newtheorem{theorem}{\indent Theorem}
\newtheorem{conjecture}{\indent Conjecture}

\noindent\textbf{CAQW based quantum hash function}

At first, we propose CAQW for constructing quantum hash function.

CAQW take place in the product space ${\mathcal{H}}_{p}\bigotimes {\mathcal{H}}_{c}$.
Let $|\ x, y,\gamma\rangle$ be a basis state, where $x,y$ and $\gamma$ represent the walker's position and the coin state respectively.

In every step of the CAQW, the evolution of the whole system can be described by the global unitary operator, denoted by $U$,
{\setlength\arraycolsep{2pt}
\begin{eqnarray}\label{EuqA01}
    U=S_y({\mathcal{I}}\otimes  C)S_x({\mathcal{I}}\otimes C).
\end{eqnarray}}The shift operator $S_x$ is defined as{\setlength\arraycolsep{2pt}
\begin{eqnarray}\label{EuqA02}
\begin{split}
    S_x=& &\left(\sum_{x\in\{1\cdots n-1\}}|\ x+1\ \rangle\langle\  x|+|\ 1\ \rangle\langle\ n|\right)\otimes |\uparrow\ \rangle\langle\ \uparrow|
    \\
        & & +\left(\sum_{x\in\{2\cdots n\}}|\ x-1\ \rangle\langle\  x|+|\ n\ \rangle\langle\ 1|\right)\otimes |\downarrow\ \rangle\langle\ \downarrow|.
\end{split}
\end{eqnarray}And $S_y$ is similar to $S_x$. The coin operator $C$ is a 2$\times$2 unitary operator.  The general coin operator is defined as
{\setlength\arraycolsep{2pt}
\begin{eqnarray}\label{EuqA03}
    C=
\left(
  \begin{array}{cccc}
    cos(\theta) & sin(\theta)  \\
    sin(\theta) & -cos(\theta)  \\
  \end{array}
\right).
\end{eqnarray}}For CAQW, we randomly select two parameters $\theta_1, \theta_2$ to construct two coin operators $C_{0}$ and $C_{1}$. The choice of the coin operator is controlled by a binary string, i.e. message. When the $n$th bit of the message is 0(1), the $n$th step of the quantum walk executes with the coin operator $C_{0}(C_{1})$. For example, if the message is 0100, the final state is expressed by
\begin{equation}
\begin{array}{l}
|\psi_{4}\rangle=U_{0}U_{0}U_{1}U_{0}|\psi_{0}\rangle,
\end{array}
\end{equation}where $|\psi_{0}\rangle$ is the initial state of the total quantum system, $U_{0}=S_y({\mathcal{I}}\otimes  C_{0})S_x({\mathcal{I}}\otimes C_{0})$, $U_{1}=S_y({\mathcal{I}}\otimes  C_{1})S_x({\mathcal{I}}\otimes C_{1})$.

Hence the probability of finding the walker at position $(x,y)$ after $t$ steps is {\setlength\arraycolsep{2pt}
\begin{eqnarray}\label{EuqA05}
    P(x,y,t)=\sum_{\gamma\in\{\uparrow,\downarrow\}}|\ \langle x,y,\gamma|U(message)|\psi_{0}\rangle\ |^2,
\end{eqnarray}}where  $U(message)$ is the global unitary operator controlled by the message.

Then, the CAQW based quantum hash function is constructed as follows.

1) Select the parameters $(n, k, (\theta_1,\theta_2), (\alpha,\beta))$. $\theta_1,\theta_2\in(0,\pi/2)$ and $|\alpha|^2+|\beta|^2=1$.

2) Run the two-dimensional CAQW under the control of the message. Each direction of the two-dimensional space is a $n$-length circle. $\theta_1,\theta_2$ are the parameters of the two coin operators respectively. The initial state is $|0,0\rangle(\alpha|0\rangle+\beta|1\rangle)$.

3) Multiply all values in the resulting probability distribution by $10^8$ modulo $2^k$ to form a binary string as the hash value. The bit length of the hash value is $n^2k$.
\vspace{3mm}

\noindent\textbf{Security of CAQW based quantum hash function}

Security of CAQW based quantum hash function is based on the infinite possibilities of the initial state and the irreversibility of measurement and modulo operator.

By using modulo operator, probability distribution is transformed to hash value. This process is irreversible because it is a many-to-one  relationship. The probability to transform the hash value back to right probability distribution is approximately 0. It is the first shield to prevent message and initial coin state from unauthorized person.

The second shield of this quantum hash function is the irreversibility of measurement. The final state of CAQW is
{\setlength\arraycolsep{2pt}
\begin{eqnarray}\label{EuqB01}
    |\psi_{t}\rangle=U(message)|\psi_{0}\rangle=\sum_{x,y} \sum_{\gamma\in\{\uparrow,\downarrow\}}\lambda_{x,y,\gamma}|\ x,y,\gamma\ \rangle.
\end{eqnarray}} This state is a pure state and is linear with the initial state. The probability distribution is
{\setlength\arraycolsep{2pt}
\begin{eqnarray}\label{EuqB02}
   P(x,y,t)=\sum_{\gamma}|\ \langle x,y,\gamma|U(message)|\psi_{0}\rangle\ |^2=\sum_{{\gamma}}|\ \lambda_{x,y,\gamma}\ |^2.
\end{eqnarray}}As a result, one can get the probability distribution easily by a quantum computer or a classical computer. However, the probability distribution is the sum of squares of amplitudes which will break the linearity between the final state and the initial state.

If an unauthorized person Eve has plaintext-cipher pairs, he still can not turn the probability distribution into the right linear composition of squares of amplitudes, let alone he doesn't have the accurate probability distribution. This is because of the infinity of decomposing a number as the sum of squares,
{\setlength\arraycolsep{2pt}
\begin{eqnarray}\label{EuqB03}
P(x,y,t)=\sum_{\gamma}|\ \lambda_{x,y,\gamma}\ |^2=\sum_{\gamma}|\ \lambda_{x,y,\gamma}^{'}\ |^2.
\end{eqnarray}}Then, Eve can only suppose the final state is
{\setlength\arraycolsep{2pt}
\begin{eqnarray}\label{EuqB04}
|\psi_{t}^{'}\rangle=\sum_{x,y} \sum_{\gamma}\lambda_{x,y,\gamma}^{'}|\ x,y,\gamma\ \rangle.
\end{eqnarray}Therefore, he can not get the right initial state even he knows the right message.
\begin{eqnarray}\label{EuqB05}
|\psi_{0}^{'}\rangle=U(message)^{-1}|\psi_{t}^{'}\rangle.
\end{eqnarray}}This process protects the initial coin state from any unauthorized person. Together with the infinite possibilities of the initial state which is aleph-one, even with a powerful quantum computer, nobody can seek out the initial state by trying all possibilities in theory.

\vspace{3mm}

\noindent\textbf{Properties of CAQW based quantum hash function}

In this section, we performed several hash tests and theoretical analysis to evaluate the performance of the proposed quantum hash function. We choose $n=5$, $k=8$, so the hash value we consider here is 200 bits. The result shows that CAQW based quantum hash function have outstanding statistical performance.
\vspace{3mm}

\noindent\textbf{Sensitivity of hash value to message}

C1,C2,C3 and C4 represent the message, and the messages with tiny modifications respectively. The results listed below show the high sensitivity to the message and the tiny changes.

Condition 1: The original message;

Condition 2: Change the 8th bit from 0 to 1;

Condition 3: Delete the last bit of the message;

Condition 4: Insert a bit in front of the 100th bit.

The corresponding 200-bit hash values in the hexadecimal format are given by:

Condition 1: F4D7DFFE8A6F9269CFF39B665D9D33FFE0912551E598438C35;

Condition 2: 9714B30709B92AC7DBC6909D95F8C5DE85F7907BD5430953E1;

Condition 3: B9CF2BA89451E17A5BC2AFEF7072A7A1AC469A644FB754B773;

Condition 4: D709199C062129047E6C68F4D5DEE1EE4E0307490A92A7CE90.

The plots of the hash values are shown respectively in Fig. 1 and it is clearly indicated that any tiny modification to the message will cause a substantial change in the final hash value. \begin{figure}[!htb]\label{Fig001}
 \begin{center}
 \caption{Plots of the 200-bit Hash Value C1, C2, C3 and C4}
  \includegraphics[width=10cm]{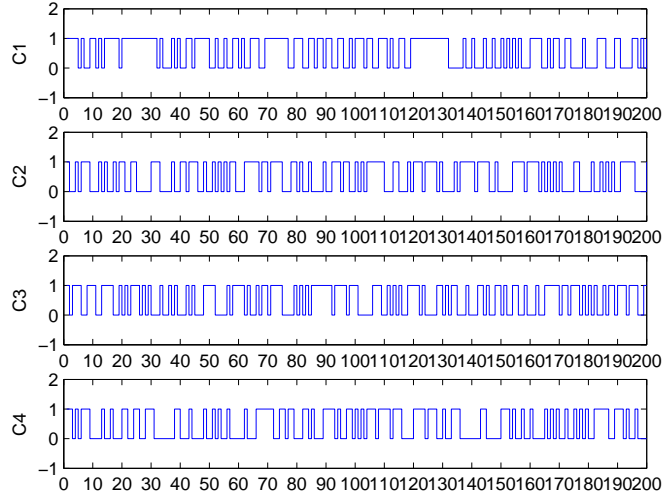}\\
  \end{center}
  \renewcommand{\figurename}{Fig.}
\end{figure}

\vspace{3mm}

\noindent\textbf{Statistical analysis of diffusion and confusion}

The diffusion and confusion tests are performed as follows:

(1) Select a message and generate the corresponding hash value;

(2) Change one bit of the message randomly and generate a new hash value;

(3) Compare the two hash values and count the changed bits called $B_i$;

(4) Repeat steps (1) to (3) $N$ times.

Given the definitions:

Minimum changed bit number $B_{min}=min(\{B_i\}^N_1)$;

Maximum changed bit number $B_{max}=max(\{B_i\}^N_1)$;

Mean changed bit number $\overline{B}=\sum^N_{i=1}B_i/N$;

Mean changed probability $P=(\overline{B}/200)\times100\%$;

Standard variance of the changed bit number $\bigtriangleup B=\sqrt{\frac{1}{N-1}\sum^N_{i=1}(B_i-\overline{B})^2}$;

Standard variance of the changed probability $\bigtriangleup P=\sqrt{\frac{1}{N-1}\sum^N_{i=1}(B_i/200-P)^2}\times100\%$.

The diffusion and confusion tests are perforemed with $N=1024$, 2048, 10000, respectively, as shown in Table \ref{Table1}. We concluded from the tests that the mean changed bit number $\overline{B}$ and the mean changed probability $P$ are close to the ideal value 100 and $50\%$ respectively. $\bigtriangleup B$ and $\bigtriangleup P$ are very little, $B_{min}$ and $B_{max}$ are around 100, so that it demonstrates the stability of diffusion and confusion. The excellent statistical effect ensures that it is impossible to forge plaintext-cipher pairs given known plaintext-cipher pairs. \begin{table}
  \centering
  \caption{Static Number of Changed Bit $B$}\label{Table1}
\begin{tabular}{|c|c|c|c|}
\hline
 & $N=1024$ & $N=2048$ & $N=10000$  \\
\hline
$\overline{B}$ &  100.1553 & 100.2036 &  99.9010  \\
\hline
$P(\%)$ &  50.0776 & 50.1018 & 49.9505 \\
\hline
$ \bigtriangleup B$ &  7.1816  &  7.0323 &  7.1133    \\
\hline
$ \bigtriangleup P$ &  3.5908  &  3.5162 &  3.5567  \\
\hline
$ B_{min} $ &  77  &  77 &  75    \\
\hline
$ B_{max}$ &  121  &  124 &  124  \\
\hline
\end{tabular}
\end{table}

\vspace{3mm}

\noindent\textbf{Collision analysis}

It is hard to provide a mathematical proof on the capability of collision resistance of chaotic hash functions. Thus, we performed the following tests for collision resistance:

(1) Select an original message randomly and generate the corresponding hash value in ASCII format.

(2) Change a bit in the message randomly and generate the corresponding hash value in ASCII format.

(3) Compare these two hash values and count the number of ASCII characters with the same value at the same location.

Moreover,  the number of ASCII characters with the same value at the same location, i.e. $\omega$, and the theoretical number of $\omega$ same values through $N$ independent tests, i.e. $W_N(\omega)$ can be computed according to the following formulas:{\setlength\arraycolsep{2pt}
\begin{eqnarray}\label{EuqC02}
&&\omega=\sum^{n^2}_{i=1}\delta(e_i-e'_i), \ \ \  where \ \ \delta(x) \ \ is \ \ the \ \ Dirac \ \ delta \ \ function.
\hspace{1mm}
\end{eqnarray}}{\setlength\arraycolsep{2pt}
\begin{eqnarray}\label{EuqC03}
&&W_N(\omega)=N\times Prob\{\omega\}=N\frac{n^2!}{\omega!(n^2-\omega)!}\left( \frac{1}{2^k}\right)^\omega \left(1- \frac{1}{2^k}\right)^{n^2-\omega}
\hspace{1mm}
\end{eqnarray}}where $e_i$ and $e'_i$ are the $i$th entries of the original and new hash values in ASCII format, respectively. In equation \ref{EuqC02}, $\omega=0,1,\cdots, n^2.$ We run this test $N=10,000$ times. The experimental values and the of experimental values $W_N(\omega)$ in the proposed function are shown in Table 2. The experimental values of $W_N(\omega)$ are similar to the theoretical values.

\begin{table}[!htb]
  \centering
  \caption{ Comparison of Experimental Values and Theoretical Values of $W_N(\omega)$}\label{Table2}
\begin{tabular}{|c|c|c|c|c|c|}
\hline
 & $\omega=0$ & $\omega=1$ & $\omega=2$  & $\omega=3$ & $\omega=4\cdots25$\\
\hline
Experimental Values of $W_N(\omega)$ &  8982 & 989 &  25 &  4 &  0  \\
\hline
\ \ Theoretical Values of $W_N(\omega)$ &  9068 & 889 &  42  &  1 &  0 \\
\hline
\end{tabular}
\end{table}

\vspace{3mm}

\noindent\textbf{Uniform distribution on hash space}

In order to check the distribution capacity in hash space, we generated two hash values according to the method described in previous subsection and then counted the number of the changed bits at each location. The statistical results for $N = 10,000$ are shown in  Fig. 2. The mean of the changed bit number 4998.2  is very close to the ideal value 5000, which accounts for half of the test times. It can be concluded that the hash value is distributed uniformly in the hash space as all the changed bit numbers are around the ideal value. Obviously, this demonstrates the resistance against statistical attack.
\begin{figure}[!htb]\label{Fig002}
 \begin{center}
 \caption{Uniform Distribution on Hash Space.}
  \includegraphics[width=10cm]{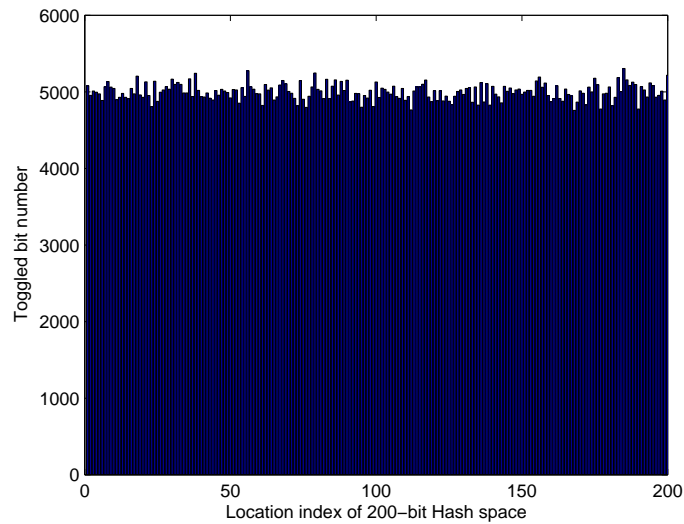}\\
  \end{center}
  \renewcommand{\figurename}{Fig.}
\end{figure}

\vspace{3mm}

\noindent\textbf{Resistance to birthday attack}

Birthday attack implies a lower bound of the length of hash value. The length of the hash value we considered here is $200=5^2\times8 $ bits. Therefore, it needs $2^{n/2}=2^{100}\approx 1.2677\times 10^{30}$ trials ($n$ is the size of hash value) to find two messages with identical hash values with a probability of 1/2. Furthermore,  CAQW based quantum hash function can be easily extended to be $392=7^2\times8$ bits or more. Therefore, the results of the tests, the size of the hash value, and the collision resistance of the proposed algorithm suggest that the birthday attack is almost impossible and that the proposed algorithm is resistant against this type of attack.

\vspace{3mm}

\noindent\textbf{Advantages}

Through the analysis above, we find that CAQW based quantum hash function has outstanding statistically performance, which is basically same with CIQW based quantum hash function. However, at the same time, CAQW based quantum hash function has some advantages than CIQW based quantum hash function due to the structure of CAQW.

In Ref. \cite{A01}, we considered two interactions: $\mathcal{I}$-interaction and $\pi$-phase interaction. $\mathcal{I}$-interaction and $\pi$-phase interaction are in fact a coin operator spanned by two $2\times2$ unitary operators. But  CIQW with these two interactions have predictable collisions that quantum walks with same special initial coin state but different messages may have the same probability distribution. In Ref. \cite{A02, A03}, to avoid this kind of collisions, the Grover operator, a swap operator\upcite{C01} and the coin operator E\upcite{C02} are chosen. But it is more hard to implement these $4\times4$ coin operators  than a coin operator spanned by two $2\times2$ unitary operators.

CAQW use the $2\times2$ coin operator in Euq. \ref{EuqA03} to control the movement of the quantum walker on two directions. By selecting two values as the parameter $\theta$, two coin operators for CAQW can be constructed. Our tests prove that predictable collisions do not happen in CAQW. The more important advantage of CAQW is that implementing two  $2\times2$ unitary operators significantly reduces the resources necessary for its feasible experimental realization than implementing a coin operator spanned by two $2\times2$ unitary operators, let alone a $4\times4$ coin operator.

A comparison about three kinds of quantum walks is shown at Table \ref{Table1}. $C_2$ and $C_4$ denote coin operators of size $2\times2$ and $4\times4$ respectively. It is obviously that CAQW is the best of them for building a quantum hash function from the aspect of safety and implementability.

\begin{table}[!hbp]
\centering
\caption{Comparison of Three Kinds of Quantum Walks}
\begin{tabular}{|c|c|c|}
\hline
CIQW in Ref. \cite{A01} & CIQW in Ref. \cite{A02, A03} & CAQW \\
\hline
Two particles &  Two particles & One particle \\
\hline
Two directions &  Two directions & Two directions \\
\hline
$S_{xy}(I\otimes C_2\otimes C_2)$ &  $S_{xy}(I\otimes C_4)$ & $S_yC_2S_xC_2$ \\
\hline
Predictable collisions &  NO & NO \\
\hline
\end{tabular}
\end{table}

Another important advantage of CAQW is that two-dimensional CAQW can be easily extended to higher dimensional CAQW. That means it is easy to construct a hash function with longer hash value. For example, when $N=5$, $k=8$, the bit length of the Hash value is $1000=5^3*8$ for three-dimensional CAQW based quantum Hash scheme, while the bit length of the Hash value is $200=5^2*8$  for two-dimensional CAQW based quantum Hash scheme.

\vspace{3mm}

\noindent\textbf{Discussion}

In this paper, based on the presentation of controlled alternative quantum walks, we introduce CAQW based quantum hash function. Security of CAQW based quantum hash function is based on the infinite possibilities of the initial state and the irreversibility of measurement and modulo operator. Furthermore, CAQW based quantum hash function has outstanding statistical  performance from the aspects of sensitivity,  diffusion and confusion, collision, uniform distribution, resistance to birthday attack. More importantly, due to the structure of alternative quantum walks, CAQW based quantum hash function has some advantages than CIQW based quantum hash functions. CAQW based quantum hash function doesn't have predictable collisions in Ref. \cite{A01}. Furthermore, implementing CAQW based quantum hash function significantly reduces the resources necessary for its feasible experimental realization than implementing CIQW based quantum hash functions in Ref. \cite{A01, A02, A03}. Also, two-dimensional CAQW based quantum hash function can be easily extended to be higher dimensional CAQW based quantum hash function, which will speed up implementing 200 bits quantum hash function. Therefore, CAQW based quantum hash function is eligible for privacy amplification in quantum key distribution, pseudo-random number generation and image encryption \cite{A03}.

\section*{Acknowledgments}
This work is supported by NSFC (Grant Nos.61571226, 61572053), the Beijing Natural Science Foundation (Grant No. 4162005), Natural Science Foundation of Jiangsu Province, China (Grant No. BK20140823), Fundamental Research Funds for the Central Universities (Grant No. NS2014096).

\noindent\textbf{Author contributions}

D. Li initiated the idea. D. Li and  J.L. Bi made the numerical simulations.  All authors reviewed the manuscript.\\
\noindent\textbf{Additional information}\\
\noindent{Competing financial interests:} The authors declare no competing financial interests.\\
\noindent\textbf{Figure legends}\\
Figure-1(Li) \emph{Plots of the 200-bit Hash Value C1, C2, C3 and C4}\\
Figure-2(Li) \emph{Uniform Distribution on Hash Space}\\
Table-1(Li) \emph{Static Number of Changed Bit B}\\
Table-2(Li) \emph{Comparison of Experimental Values and Theoretical Values of $W_N(\omega)$}\\
Table-3(Li) \emph{Comparison of Three Kinds of Quantum Walks}\\


\begin{thebibliography}{}

\bibitem{A04} Bennett, C.H., Brassard, G., Cr\'{e}peau, C. \textbf{\&} Maurer, U.M. Generalized privacy amplification. \emph{IEEE Trans. Inf. Theory}  \textbf{41} 1915-1923 (1995).

\bibitem{A01} Li, D., Zhang, J., Guo, F.Z., Huang, W., Wen, Q.Y. \textbf{\&} Chen, H. Discrete-time interacting quantum walks and quantum hash schemes.  \emph{Quant. Inf. Proc.} \textbf{3} 1501-1513 (2013).

\bibitem{A02} Li, D., Zhang, J., Ma, X. W., Zhang, W. W. \textbf{\&} Wen, Q.Y. Analysis of the two-particle controlled interacting quantum walks. \emph{Quant. Inf. Proc.}  \textbf{6}  2167-2176 (2013).

\bibitem{A03} Yang, Y.G., Xu, P., Yang, R., Zhou, Y.H. \textbf{\&} Shi, W.M. Quantum hash function and its application to privacy amplification in quantum key distribution, pseudo-random number generation and image encryption.  \emph{Sci. Rep.}  \textbf{6} 19788 (2016).

\bibitem{A10} El\'{\i}as, S. \textbf{\&} Andraca, V. Quantum walks: a comprehensive review. \emph{Quant. Inf. Proc.} \textbf{11} 1015-1106 (2012).

\bibitem{A11}  Ambainis, A. Quantum walk algorithm for element distinctness. \emph{SIAM J. Comput.} \textbf{37} 210 (2007).

\bibitem{A12}  Shenvi, N., Kempe, J. \textbf{\&} Whaley, B.K. Quantum random-walk search algorithm. \emph{Phys. Rev. A} \textbf{67} 052307 (2003).

\bibitem{A13}  Berry, S.D., Wang, J.B. Two-particle quantum walks: Entanglement and graph isomorphism testing. \emph{Phys. Rev. A} \textbf{83} 042317 (2011).

\bibitem{A14} Zhan, X., Qin, H., Bian, Z.H., et al. Perfect state transfer and efficient quantum routing: A discrete-time quantum-walk approach. \emph{Phys. Rev. A} \textbf{90} 012331 (2014).

\bibitem{A15} Babatunde, A.M., Cresser, J. \textbf{\&} Twamley, J. Using a biased quantum random walk as a quantum lumped element router.  \emph{Phys. Rev. A} \textbf{90} 012339 (2014).

\bibitem{A16} Kurzynski, P. \textbf{\&} Wojcik, A. Quantum walk as a generalized measuring device. \emph{Phys. Rev. L} \textbf{110} 200404 (2013).

\bibitem{B01} Di Franco, C.,  Mc Gettrick, M., \textbf{\&} Th. Busch. Mimicking the Probability Distribution of a Two-Dimensional GroverWalk with a Single-Qubit Coin. \emph{Phys. Rev. Lett} \textbf{106} 080502 (2011).

\bibitem{B02} Di Franco, C.,  Mc Gettrick, M., Machida, T., \textbf{\&} Th. Busch. Alternate two-dimensional quantum walk with a single-qubit coin. \emph{Phys. Rev. A} \textbf{84} 042337 (2011).

\bibitem{B03} Roldan, E., Di Franco, C., Silva, F., de Valcarcel, G.J. N-dimensional alternate coined quantum walks from a dispersion-relation perspective. \emph{Phys. Rev. A} \textbf{87}, 022336 (2013).

\bibitem{B04}  Di Franco, C.,  Mc Gettrick, M., Machida, T., \textbf{\&} Th. Busch. Measurement-induced generation of spatial entanglement in a two-dimensional quantum walk with single-qubit coin. \emph{Journal of Computational \textbf{\&} Theoretical Nanoscience}  \textbf{7} 1613 (2013).

\bibitem{B05} Di Franco, C. and  Paternostro, M. Localizationlike effect in two-dimensional alternate quantum walks with periodic coin operations. \emph{Phys. Rev. A} \textbf{91} 012328 (2015).

\bibitem{B06} Machida, T., Chandrashekar, C. M. Localization and limit laws of a three-state alternate quantum walk on a two-dimensional lattice. \emph{Phys. Rev. A} \textbf{92} 062307 (2015).

\bibitem{B07} Bru, L.A., de Valcarcel, G.J., Di Molfetta, G., Perez, A., Roldan, E. \textbf{\&} Silva, F. Quantum walks on  a cylinder. \emph{Phys. Rev. A} \textbf{94} 032328 (2016).

\bibitem{B08} Chen, T., Zhang, X.D. Extraordinary behaviors in a two-dimensional decoherent alternative quantum walk. \emph{Phys. Rev. A} \textbf{94} 012316 (2016).

\bibitem{C01} Xue, P., Sanders, B.C. Two quantum walkers sharing coins. \emph{Phys. Rev. A} \textbf{85} 022307 (2012)

\bibitem{C02} $\check{S}$tefa$\check{n}$$\acute{a}$k, M., Barnett, S.M., Koll$\acute{a}$r, B., Kiss, T., Jex, I. Directional correlations in quantum walks with two particles. \emph{New J. Phys.} \textbf{13} 033029 (2011).

\bibitem{D} Zhang, J., Wang, X. \textbf{\&} Zhang, W. CHaotic keyed hash function based on feed forward-feedback nonlinear digital filter. \emph{Phys. Lett. A} \textbf{362} 439-448 (2007).


\end{thebibliography}
\end{document}